\documentclass[%
a4paper,
twocolumn,
twoside,
superscriptaddress,
nofootinbib,
 amsmath,amssymb,
]{revtex4-1}

\usepackage{graphicx}% Include figure files
\usepackage{dcolumn}% Align table columns on decimal point
\usepackage{bm}% bold math
\usepackage{cases}
\usepackage[hidelinks=true]{hyperref}% add hypertext capabilities
\usepackage{comment}
\begin{document}
%\linenumbers

%\preprint{APS/123-QED}

%\title{Encoding a causal structure into a neural network}
%\title{Mapping classical correlations in causal structures using neural networks}
%A neural network oracle for classical correlations on causal structures}% Force line breaks with \\
\title{A neural network oracle for quantum nonlocality problems in networks}

\author{Tam\'as Kriv\'achy\footnote{Corresponding author. \url{tamas.krivachy@gmail.com}}}
\author{Yu Cai}
\affiliation{%
Department of Applied Physics, University of Geneva, 
CH-1211 Geneva, Switzerland
}%
\author{Daniel Cavalcanti}
\affiliation{ICFO, The Institute of Photonic Sciences, 08860 Castelldefels (Barcelona), Spain}
\author{Arash Tavakoli}
\affiliation{Dyson School of Design Engineering, Imperial College London, London SW7 2AZ, UK}
\author{Nicolas Gisin}
\author{Nicolas Brunner}
\affiliation{%
Department of Applied Physics, University of Geneva, 
CH-1211 Geneva, Switzerland
}%

\date{\today}% It is always \today, today,

\begin{abstract}
Characterizing quantum nonlocality in networks is a challenging, but important problem. Using quantum sources one can achieve distributions which are unattainable classically. A key point in investigations is to decide whether an observed probability distribution can be reproduced using only classical resources. This causal inference task is challenging even for simple networks, both analytically and using standard numerical techniques. We propose to use neural networks as numerical tools to overcome these challenges, by learning the classical strategies required to reproduce a distribution. As such, the neural network acts as an oracle, demonstrating that a behavior is classical if it can be learned. We apply our method to several examples in the triangle configuration. After demonstrating that the method is consistent with previously known results, we give solid evidence that the distribution presented in [N. Gisin, Entropy 21(3), 325 (2019)] is indeed nonlocal as conjectured. Finally we examine the genuinely nonlocal distribution presented in [M.-O. Renou \emph{et al.}, PRL 123, 140401 (2019)], and, guided by the findings of the neural network, conjecture nonlocality in a new range of parameters in these distributions. The method allows us to get an estimate on the noise robustness of all examined distributions.
\end{abstract}

%Use showkeys class option if keyword
\maketitle

%\tableofcontents
\section{Introduction}
The possibility of creating stronger than classical correlations between distant parties has deep implications for both the foundations and applications of quantum theory. These ideas have been initiated by Bell~\cite{bell_einstein_1964}, with subsequesnt research leading to the theory of Bell nonlocality~\cite{brunner_bell_2014}. In the Bell scenario multiple parties jointly share a single classical or quantum source, often referred to as local and nonlocal sources, respectively. Recently, interest in more decentralized causal structures, in which several independent sources are shared among the parties over a network, has been on the rise~\cite{branciard_characterizing_2010,branciard_bilocal_2012,fritz_beyond_2012, pusey_viewpoint_2019}. Contrary to the Bell scenario, in even slightly more complex networks the boundary between local and nonlocal correlations becomes nonlinear and the local set non-convex, greatly perplexing rigorous analysis. Though some progress has been made~\cite{chaves_causal_2014, chaves_inferring_2014, henson_theory-independent_2014,tavakoli_nonlocal_2014, chaves_informationtheoretic_2015, wolfe_inflation_2019, rosset_nonlinear_2016, navascues_inflation_2017,rosset_universal_2017, chaves_polynomial_2016,fraser_causal_2018, weilenmann_non-shannon_2018, luo_computationally_2018, renou_genuine_2019, gisin_constraints_2020, renou_limits_2019, pozas-kerstjens_bounding_2019}, we still lack a robust set of tools to investigate generic networks from an analytic and numerical perspective.

Here we explore the use of machine learning in these problems. In particular we tackle the membership problem for causal structures, i.e. given a network and a distribution over the observed outputs, we must decide whether it could have been produced by using exclusively local resources. We encode the causal structure into a neural network and ask the network to reproduce the target distribution. By doing so, we approximate the question ``does a local causal model \emph{exist}?'' with ``is a local causal model \emph{learnable}?''. Neural networks have proven to be useful ans\"atze for generic nonlinear functions in terms of expressivity, ease of learning and robustness, both in- and outside the domain of physical sciences~\cite{melko_restricted_2019,iten_discovering_2020,melnikov_active_2018, van_nieuwenburg_learning_2017,carrasquilla_machine_2017}.
Machine learning has also been used in the study of nonlocality~\cite{deng_machine_2018,canabarro_machine_2019}. However, while the techniques of Ref.~\cite{canabarro_machine_2019} can only suggest if a distribution is local or nonlocal, the method employed here is generative and provides a certificate that a distribution is local once it is learned.
\begin{figure}[t!]
    \centering
    \includegraphics[width = 0.4 \textwidth]{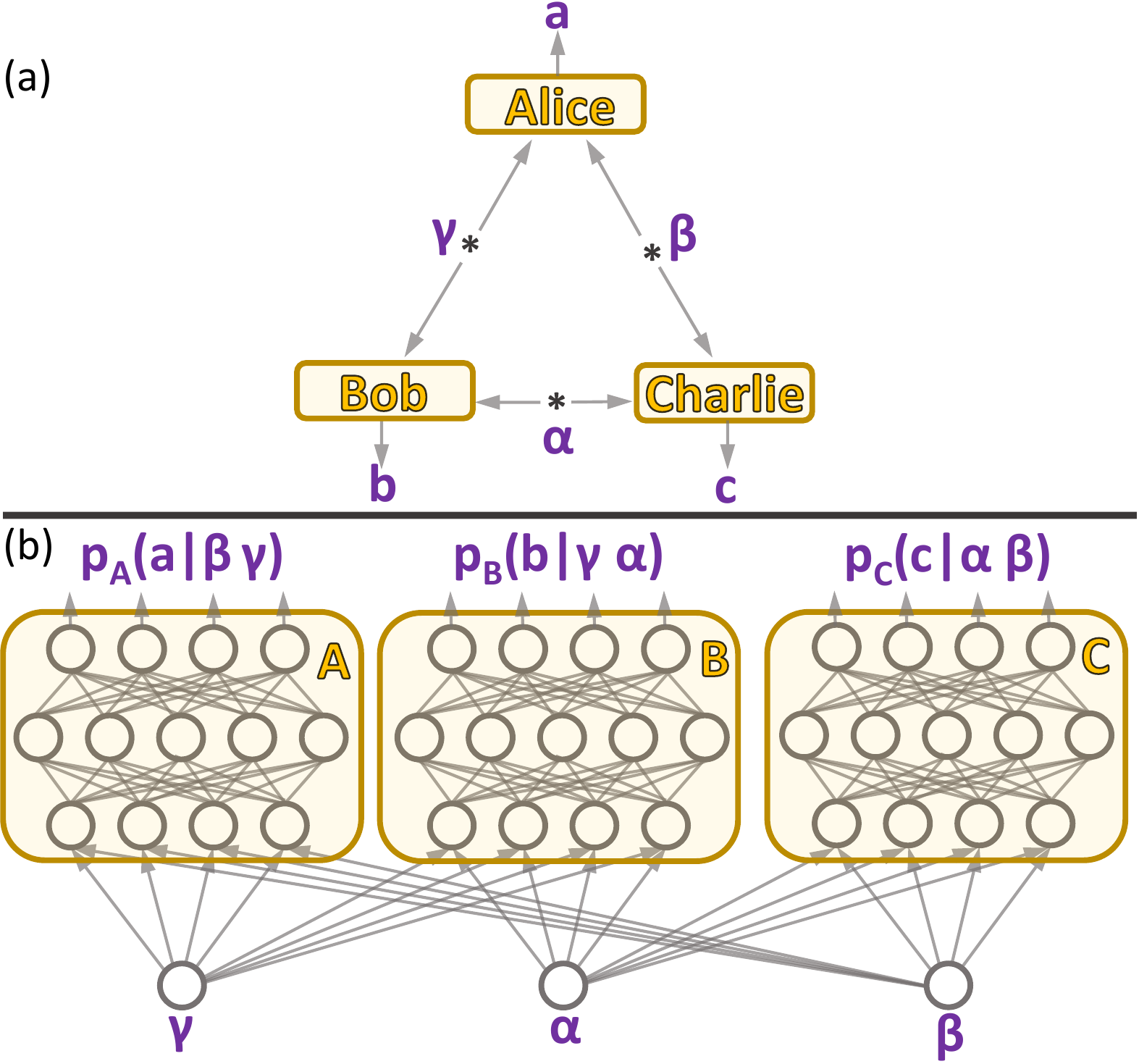}
    \caption{(a) Triangle network configuration. (b) Neural network which reproduces distributions compatible with the triangle configuration.}
    \label{fig:triangle}
\end{figure}

In our approach we exploit that both causal structures and feedforward neural networks have their information flow determined by a directed acyclic graph. For any given distribution over observed variables and an ansatz causal structure, we train a neural network which respects that causal structure to reproduce the target distribution. This is equivalent to having a neural network learn the local responses of the parties to their inputs. If the target distribution is inside the local set, then a sufficiently expressive neural network should be able to learn the appropriate response functions and reproduce it. For distributions outside the local set, we should see that the machine can not approximate the given target. This gives us a criterion for deciding whether a target distribution is inside the local set or not. In particular, if a given distribution is truly outside the local set, then by adding noise in a physically relevant way we should see a clear transition in the machine's behavior when entering the set of local correlations.

We explore the strength of this method by examining a notorious causal structure, the so-called `triangle' network, depicted in Fig.~\ref{fig:triangle}. The triangle configuration is among the simplest tripartite networks, yet it poses immense challenges theoretically and numerically. We use the triangle with quaternary  outcomes as a test-bed for our neural network oracle. After checking for the consistency of our method with known results, we examine the so-called Elegant distribution, proposed in~\cite{gisin_entanglement_2019}, and the distribution proposed by Renou~\emph{et al.} in~\cite{renou_genuine_2019}. Our method gives solid evidence that the Elegant distribution is outside the local set, as originally conjectured. The family of distributions proposed by Renou~\emph{et al.} was shown to be nonlocal in a certain regime of parameters. When examining the full range of parameters we not only recover the nonlocality in the already known regime, but also get a conjecture of nonlocality from the machine in another range of the parameters. Finally, we use our method to get estimates of the noise robustness of these nonlocal distributions, and to gain insight into the learned strategies.

\section{Encoding causal structures into neural networks}
The methods developed in this work are in principle applicable to any causal structure. Here we demonstrate how to encode a network nonlocality configuration into a neural network on the highly non-trivial example of the triangle network with quaternary  outputs and no inputs. In this scenario three sources, $\alpha,\beta,\gamma$, send information through either a classical or a quantum channel to three parties, Alice, Bob and Charlie. Flow of information is constrained such that the sources are independent from each other, and each one only sends information to two parties of the three, as depicted in Fig.~\ref{fig:triangle}. Alice, Bob and Charlie process their inputs with arbitrary local response functions, and they each output a number $a,b,c\in\{0,1,2,3\}$, respectively. Under the assumption that each source is independent and identically distributed from round to round, and that the local response functions are fixed (though possibly stochastic), such a scenario is well characterized by the probability distribution $p(abc)$ over the random variables of the outputs.

If quantum channels are permitted from the sources to the parties then the set of distributions is larger than that achievable classically. Due to the nonlocal nature of quantum theory, these correlations are often referred to as nonlocal ones, as opposed to local behaviors arising from only using classical channels. In the classical case, the scenario is equivalent to a causal structure, otherwise known as a Bayesian network~\cite{pearl_causality_2000,koller_probabilistic_2009}.

For the classical setup we can assume without loss of generality that the sources each send a random variable drawn from a uniform distribution on the continuous interval between $0$ and $1$. Given the network constraint, the probability distribution over the parties' outputs can be written as
\begin{align}\label{eq:triangle-integral}
    p(abc) = \int_{0}^1 d\alpha d\beta d\gamma \, p_A(a|\beta \gamma) p_B(b|\gamma \alpha) p_C(c|\alpha \beta).
\end{align}

We now construct a neural network which is able to approximate a distribution of the form (\ref{eq:triangle-integral}). We use a feedforward neural network, since it is described by a directed acyclic graph, similarly to a causal structure~\cite{goodfellow_deep_2016, pearl_causality_2000, koller_probabilistic_2009}. This allows for a seamless transfer from the causal structure to the neural network model. The inputs are the hidden variables, i.e. uniformly drawn random numbers $\alpha,\beta,\gamma$. The outputs are the conditional probabilities $p_A(a|\beta \gamma), p_B(b|\gamma \alpha)$ and $p_C(c|\alpha \beta)$, i.e. three normalized vectors, each of length 4. So as to respect the communication constraints of the triangle, the neural network is not fully connected, as shown in Fig.~\ref{fig:triangle}. We evaluate the neural network for $N_{batch}$ values of $\alpha,\beta,\gamma$ in order to approximate the joint probability distribution (\ref{eq:triangle-integral}) with a Monte Carlo approximation,

\begin{align}
\label{eq:triangle-sum-long}
    p_M(abc) = \frac{1}{N_{batch}}\sum_{i=1}^{N_{batch}} p_A(a|\beta_i \gamma_i) p_B(b|\gamma_i \alpha_i) p_C(c|\alpha_i \beta_i).
\end{align}
In our implementation each of the three conditional probability functions is modeled by a multilayer perceptron, with rectified linear or tangent hyperbolic activations, except at the last layer, where we have a softmax layer to impose normalization. Note, however, that any feedforward network can be used to model these conditional probabilities. The cost function can be any measure of discrepancy between the target distribution $p_t$ and the neural network's output $p_M$, such as the Kullback--Leibler divergence\footnote{We observed that for many target distributions our implementation worked well also when using the mean squared error or mean absolute error. However, the Kullback--Leibler divergence worked well with all examined distributions.} of one relative to the other, namely $\sum_{abc}p_t(abc)\log\left( \frac{p_t(abc)}{p_M(abc)}\right)$. In order to train the neural network we synthetically generate uniform random numbers for the hidden variables, the inputs. We then adjust the weights of the network after evaluating the cost function on a minibatch of size $N_{batch}$, using conventional neural network optimization methods~\cite{goodfellow_deep_2016}. The minibatch size is chosen arbitrarily and can be increased in order to increase the neural network's precision. For the triangle with quaternary outputs an $N_{batch}$ of several thousands is typically satisfactory.

By encoding the causal structure in a neural network like this, we can train the neural network to try to reproduce a given target distribution. The procedure generalizes in a straight-forward manner to any causal structure, and is thus in principle applicable to any quantum nonlocality network problem. We provide specific code online for the triangle configuration, as well as for the standard Bell scenario, which has inputs as well (see Section~\ref{sec:code}). After finishing this work we realized that related ideas have been investigated in causal inference, though in a different context, where network architectures and weights are simultaneously optimized to reproduce a given target distribution over continuous outputs, as opposed to discrete ones examined here~\cite{goudet_learning_2018}.

\section{Results}
\begin{figure}[b!]
    \centering
    \includegraphics[width = 0.44 \textwidth]{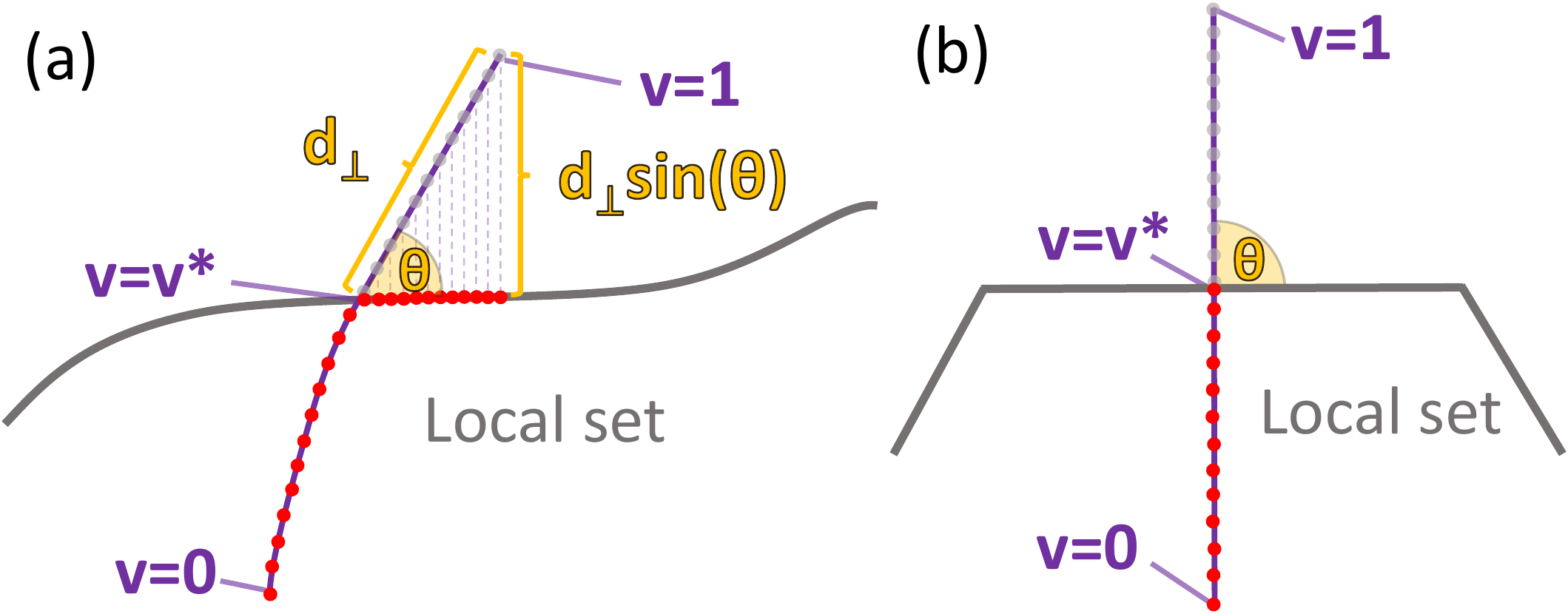}
    \caption{Visualization of target distributions $p_t(v)$ leaving the local set at an angle $\theta$ for a generic noisy distribution (a) and for the specific case of the Fritz distribution with a 2-qubit Werner state shared between Alice and Bob (b). The grey dots depict the target distributions, while the red dots depict the distributions which the neural network would find. In the generic case we depict the distance $d_\bot:=d(p_t(v),p_t(v^*))$ introduced in Eq.~\ref{eq:true-distance}, for the special case of $v=1$, as well as $d_\bot\sin\theta$. Given an estimate for $v^*$, the distance $d_\perp$ can be evaluated analytically, which (for an appropriate $\theta$) allows us to compare $d_\bot \sin \theta$ with the distance that the machine perceives.}
    \label{fig:leaving-local-set-1}
\end{figure}
\begin{figure*}[t!]
    \centering
    \includegraphics[width = \textwidth]{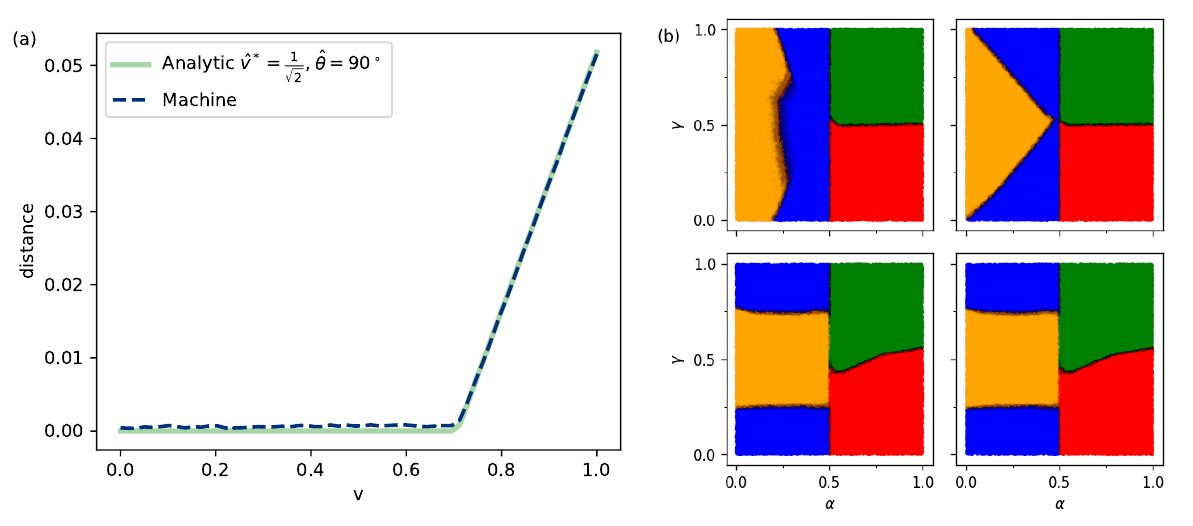}
    \caption{Fritz distribution~\cite{fritz_beyond_2012} results. (a) Plot of the distance perceived by the machine, $d_M(v)$ and the analytic distance $\hat d(v)$ for $\hat v^*=1/\sqrt{2}$ and $\hat \theta = 90^\circ$. (b) Visualization of response functions of Bob as a function of $\alpha, \gamma$ for $v=0, 0.44, 0.71, 1$, from top left to bottom right, respectively. Note how the responses for $v>\hat v^*$ are the same.}
    \label{fig:Fritz}
\end{figure*}

Given a target distribution $p_t$, the neural network provides an explicit model for a distribution $p_M$, which is, according to the machine, the closest local distribution to $p_t$. The distribution $p_M$ is guaranteed to be from the local set by construction. The neural network will almost never exactly reproduce the target distribution since $p_M$ is learned by sampling a distribution a finite number of times, and additionally the learning techniques do not guarantee convergence to the global optimum. As such, to use the neural network as an oracle we could define some confidence level for the similarity between $p_M$ and $p_t$. It is, however, more robust and informative if instead, we search for transitions in the machine's behavior when giving it different target distributions from both outside the local set and inside it. We will typically define a family of target distributions $p_t(v)$ by taking a distribution which is believed to be nonlocal and adding some noise controlled by the parameter $v$, with $p_t(v=0)$ being the completely noisy (local) distribution and $p_t(v=1)$ being the noiseless, ``most nonlocal'' one. By adding noise in a physically meaningful way we guarantee that at some parameter value, $v^*$, we will enter the local set and stay in it for $v<v^*$. For each noisy target distribution we retrain the neural network and obtain a family of learned distributions $p_M(v)$. Observing a qualitative change in the machine's performance at some point is an indication of traversing the local set's boundary. In this work we extract information from the learned model through
\begin{itemize}
\item the distance between the target and the learned distribution,
\[d(p_t,p_M) = \sqrt{\sum_{abc} \left[p_t(abc)-p_M(abc)\right]^2,}\]
\item the learned distributions $p_M(v)$, in particular by examining the local response functions of Alice, Bob and Charlie.
\end{itemize}

Observing a clear liftoff of the distance $d_M(v) := d(p_t(v),p_M(v))$ at some point is a signal that we are leaving the local set. Somewhat surprisingly, we can deduce even more from the distance $d_M(v)$. Though the shape of the local set and the threshold value $v^*$ are unknown, in some cases, under mild assumptions, we can estimate not only $v^*$, but also the angle at which the curve $p_t(v)$ exits the local set, and in addition gain some insight into the shape of the local set near $p_t(v^*)$. To do this, let us first assume that the local set is flat near $p_t(v^*)$ and that $p_t(v)$ is a straight curve. Then the true distance from the local set is 

\begin{align}
\label{eq:true-distance}
d(v) = 
  \begin{cases}
   0        & \text{if } v \leq v^* \\
   d\left(p_t(v),p_t(v^*)\right) \sin(\theta)      & \text{if } v > v^*,
  \end{cases}
\end{align}
where $\theta$ is the angle between the curve $p_t(v)$ and the local set's hyperplane (see Fig.~\ref{fig:leaving-local-set-1} for an illustration). In the more general setting Eq.~(\ref{eq:true-distance}) is still approximately correct even for $v>v^*$, if $p_t(v)$ is almost straight and the local set is almost flat near $v^*$. We denote this analytic approximation of the true distance form the local set as $\hat{d}(v)$. We use Eq.~(\ref{eq:true-distance}) to calculate it but keep in mind that it is only an approximation. Given an estimate for the two parameters $v^*$ and $\theta$ this function can be compared to what the machine perceives as a distance, $d_M(v)$. Finding a match between the two distance functions gives us strong evidence that indeed the curve $p_t(v)$ exits the local set at $\hat v^*$ at an angle $\hat\theta$, where the hat is used to signify the obtained estimates.

We also get information out of the learned model by looking at the local responses of Alice, Bob and Charlie. Recall that the shared random variables, the sources, are uniformly distributed, hence the response functions encode the whole problem. We can visualize, for example, Bob's response function $p_B(b|\alpha,\gamma)$ by sampling several thousand values of $\{\alpha,\gamma\}\in [0,1]^2$. In order to capture the stochastic nature of the responses, for each pair $\alpha,\gamma$ we sample from $p_B(b|\alpha,\gamma)$ thirty times and color-code the results $b\in\{$red, blue, green, yellow$\}$. By scatter plotting these points with a finite opacity we gain an impression of the response function, such as in Fig.~\ref{fig:Fritz}.

These figures are already interesting in themselves and can guide us towards analytic guesses of the ideal response functions. However, they can also be used to verify our results in some special cases. For example, if $\theta = 90^\circ$ and the local set is sufficiently flat, then the response functions should be the same\footnote{For any target distribution the closest local response function is not unique, so response functions could vary above $v^*$. However after running the algorithm for the full range of $v$, for each $v$ we check whether the models at other $v'$ values perform better for $p_t(v)$. This smooths the results and gives more consistent response functions.} for all $v\geq v^*$, as it is in Fig.~\ref{fig:Fritz}. On the other hand if $\theta < 90^\circ$ then we are in a scenario similar to that of the Fig.~\ref{fig:leaving-local-set-1}(a) and the response functions should differ for different values of $v$.

\begin{figure*}[t!]
    \centering
    \includegraphics[width = \textwidth]{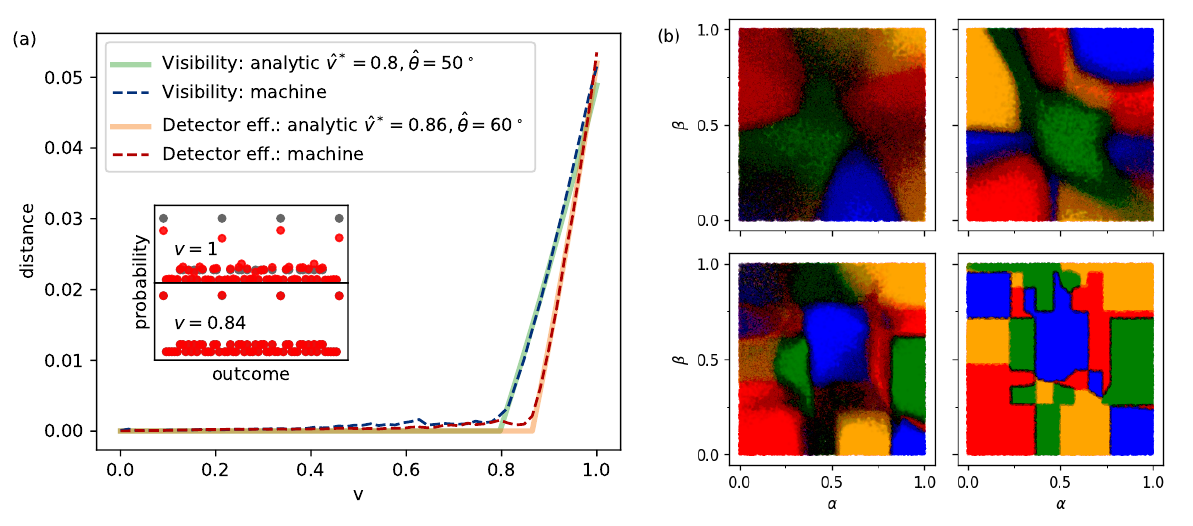}
    \caption{Elegant distribution~\cite{gisin_entanglement_2019} results. (a) Comparison of the distance perceived by the machine, $d_M(v)$ and the analytic distance $\hat d(v)$. Both visibility and detector efficiency model results are shown. Inset: The target (gray) and learned (red) distributions visualized by plotting the probability of each of the 64 possible outcomes, for detector efficiency $v=1$ and $v=0.84$. Note that for $v=0.84$ most gray dots are almost fully covered by the corresponding red dots. (b) Responses of Charlie illustrated as a function of $\alpha,\beta$. Detector efficiency values (top left to bottom right): $v =$ 0.5, 0.72, 0.76, 1.}
    \label{fig:elegant}
\end{figure*}

\subsection{Fritz distribution}
First let us consider the quantum distribution proposed by Fritz~\cite{fritz_beyond_2012}, which can be viewed as a Bell scenario wrapped into the triangle topology. Alice and Bob share a singlet, i.e. $|\psi\rangle_{AB} = |\psi^-\rangle = \frac{1}{\sqrt{2}}\left(|01\rangle - |10\rangle\right)$, while Bob and Charlie share either a maximally entangled or a classically correlated state with Charlie, such as $\rho_{BC} = \frac{1}{2} (|00\rangle \langle 00|+ |11\rangle \langle 11|)$ and similarly for $\rho_{AC}$. Alice measures the shared state with Charlie in the computational basis and, depending on this random bit, she measures either the Pauli $X$ or $Z$ observable. Bob does the same with his shared state with Charlie and measures either $\frac{X+Z}{\sqrt{2}}$ or $\frac{X-Z}{\sqrt{2}}$. They then both output the measurement result and the bit which they used to decide the measurement. Charlie measures both sources in the computational basis and announces the two bits. As a noise model we introduce a finite visibility for the singlet shared by Alice and Bob, thus we  examine a Werner state,

\begin{align}
\label{Werner}
\rho(v) = v |\psi^-\rangle\langle \psi^-| + (1-v) \frac{\mathbb{I} }{4},
\end{align}
where $\mathbb{I}/4$ denotes the maximally mixed state of two qubits. For such a state we expect to find a local model below the threshold of $v^*=\frac{1}{\sqrt{2}}$.

In Fig.~\ref{fig:Fritz} we plot the learned ($d_M(v)$) and analytic ($\hat{d}(v)$) distances discussed previously for $\hat \theta=90^\circ$ and $\hat v^*=\frac{1}{\sqrt{2}}$. The coincidence of the two curves is already good evidence that the machine finds the closest local distributions to the target distributions. Upon examining the response functions of Alice, Bob and Charlie, also in Fig.~\ref{fig:Fritz}, we see that they do not change above $\hat v^*$, which means that the machine finds the same distributions for target distributions outside the local set. This is in line with our expectations. Due to the connection with the standard Bell scenario (where the local set is a polytope), we believe the curve $p_t(v)$ exits the local set perpendicularly, as it is depicted in Fig.~\ref{fig:leaving-local-set-1}(b). These results confirm that our algorithm functions well.

\subsection{Elegant distribution}
Next we turn our attention to a distribution which is more native to the triangle structure, as it combines entangled states and entangled measurements. We examine the Elegant distribution, which is conjectured in~\cite{gisin_entanglement_2019} to be outside the local set. The three parties share singlets and each perform a measurement on their two qubits, the eigenstates of which are

\begin{align}
|\Phi_j\rangle = \sqrt{\frac{3}{2}} |m_j, - m_j \rangle + i \frac{\sqrt{3}-1}{2} |\psi^-\rangle,
\end{align}
where the $|m_j\rangle$ are the pure qubit states with unit length Bloch vectors pointing at the four vertices of the tetrahedron for $j=1,2,3,4$, and $|- m_j\rangle$ are the same for the inverted tetrahedron.

We examine two noise models - one at the sources and one at the detectors. First we introduce a visibility to the singlets such that all three shared quantum states have the form (\ref{Werner}). Second, we examine detector efficiency, in which each detector defaults independently with probability $1-v$ and gives a random output as a result. This is equivalent to adding white noise to the quantum measurements performed by the parties, i.e. the positive operator-valued measure elements are $\mathcal{M}_j = v|\Phi_j\rangle \langle \Phi_j| + (1-v)\frac{\mathbb{I} }{4}$.

For both noise models we see a transition in the distance $d_M(v)$, depicted in Fig.~\ref{fig:elegant}, giving us strong evidence that the conjectured distribution is indeed nonlocal. Through this examination we gain insight into the noise robustness of the Elegant distribution as well. It seems that for visibilities above $\hat v^* \approx 0.80$, or for detector efficiency above $\hat v^* \approx 0.86$, the distribution is still nonlocal. The curves exit the local set at approximately $\hat \theta \approx 50^\circ$ and $\hat \theta \approx 60^\circ$, respectively. Note that for both distribution families, by looking at the unit tangent vector, one can analytically verify that the curves are almost straight for values of $v$ above the observed threshold. This gives us even more confidence that it is legitimate to use the analytic distance $\hat{d}(v)$ as a reference (see Eq.~(\ref{eq:true-distance})). In Fig.~\ref{fig:elegant} we illustrate how the response function of Charlie changes when adding detector efficiency. It is peculiar how the machine often prefers horizontal and vertical separations of the latent variable space, with very clean, deterministic responses, similarly to how we would do it intuitively, especially for noiseless target distributions.

\begin{comment}
\begin{figure}[b!]
    \centering
    \includegraphics[width = 0.5 \textwidth]{elegant-distances-joint.png}
    \caption{Comparison of the distance perceived by the machine, $d_M(v)$ and the approximation of the distance $\hat d(v)$ for the Elegant distribution. Both visibility and detector efficiency model results are shown.}
    \label{fig:elegant}
\end{figure}

\begin{figure}[h!]
    \centering
    %\includegraphics[scale=0.67]{Elegant-responses-vis.png}
    % (Top) Visibility values (top left to bottom right): 0.17, 0.68, 0.81, 1. (Bottom)
    \includegraphics[width = 0.5 \textwidth]{Elegant-responses-Charlie-1.png}
    \caption{Responses of Charlie for the Elegant distribution illustrated as a function of $\beta$, $\gamma$, similarly to Fig.~\ref{fig:Fritz}. Detector efficiency values (top left to bottom right): 0.5, 0.72, 0.76, 1.}
    \label{fig:elegant-responses}
\end{figure}
\end{comment}

\begin{figure*}[t!]
    \centering
    \includegraphics[width = \textwidth]{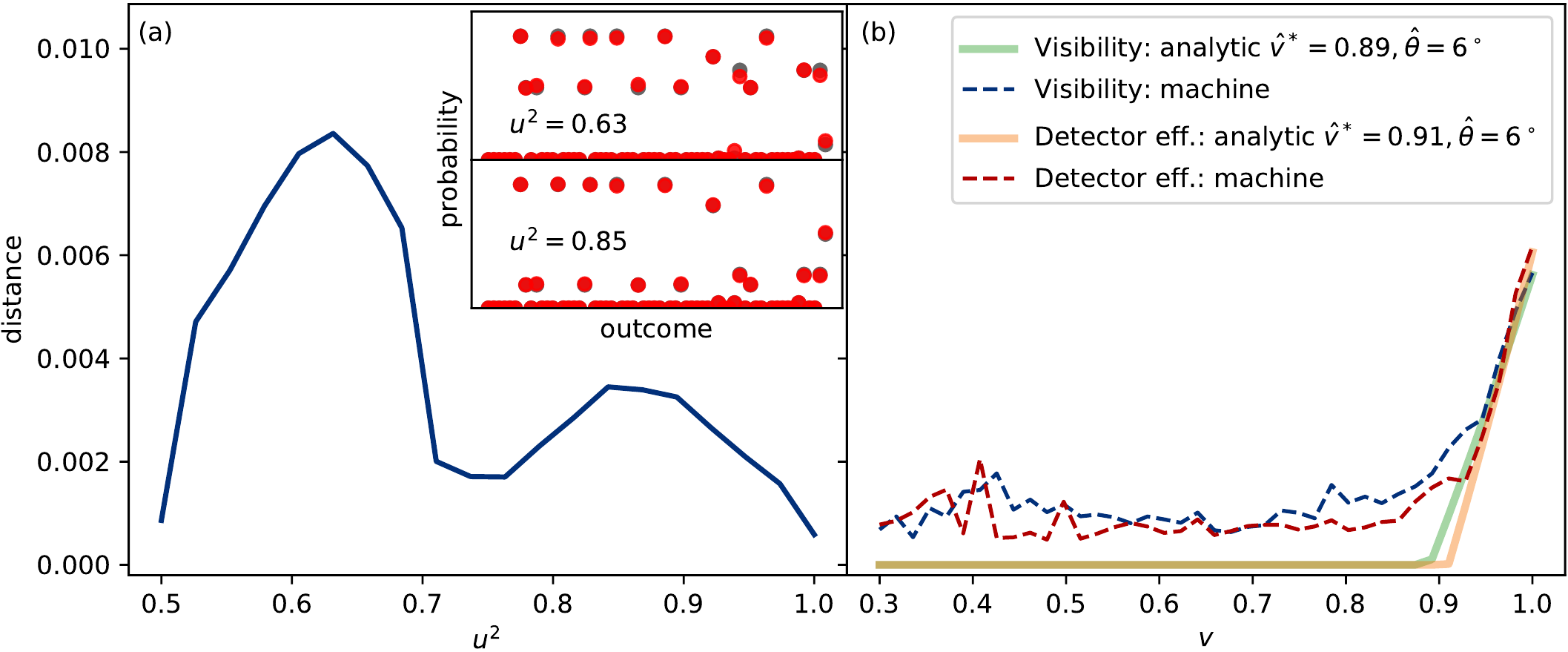}
    \caption{Renou \emph{et al.} distribution~\cite{renou_genuine_2019} results. (a) The distance perceived by the machine, $d_M$, as a function of $u^2$, with no added noise. Inset: The target (gray) and learned (red) distributions visualized by plotting the probability of each of the 64 possible outcomes, for $u^2=0.63$ and $u^2=0.85$. These $u^2$ values approximately correspond to the two peaks in the scan. Note that most gray dots are almost fully covered by the corresponding red dots. (b) Noise scans, i.e. the analytic $\hat d(v)$ (see Eq.~(\ref{eq:true-distance})) and the learned $d_M(v)$, for the target distribution of $u^2=0.85$, for the detector efficiency and visibility noise models.}
    \label{fig:Renou}
\end{figure*}
\subsection{Renou \emph{et al.} distribution}
The authors of~\cite{renou_genuine_2019} recently introduced the first distribution in the triangle scenario which is not directly inspired by the Bell scenario and is proven to be nonlocal. To generate the distribution take all three shared states to be the entangled states $|\phi^+ \rangle = \frac{1}{\sqrt{2}}\left( |00\rangle + |11\rangle \right)$. Each party performs the same measurement, characterized by a single parameter $u\in [\frac{1}{\sqrt{2}},1]$, with eigenstates $|01\rangle, |10\rangle, u|00\rangle + \sqrt{1-u^2}|11\rangle, \sqrt{1-u^2}|00\rangle - u|11\rangle$. The authors prove that for $u_{max}^2<u^2<1$ this distribution is nonlocal, where $u_{max}^2 \approx0.785$ and also show that there exist local models for $u^2 \in \{0.5,u_{max}^2,1\}$. Though they argue that there must be some noise tolerance of the distribution, they lack a proper estimation of it.

First we examine these distributions as a function of $u^2$, \emph{without} any added noise. The results are depicted in Fig.~\ref{fig:Renou}(a). To start with, note how the distances are numerically much smaller than in the previous examples, i.e. the machine finds distributions which are extremely close to the targets. See the inset in Fig.~\ref{fig:Renou} for examples which exhibit how close the learned distributions are to the targets even for the points which have large distances ($u^2 = 0.63,0.85$). We observe, consistently with analytic findings, that for $u_{max}^2<u^2<1$, the machine finds a non-zero distance from the local set. It also recovers the local models at $u^2 \in \{0.5,u_{max}^2,1\}$, with minor difficulties around $u_{max}^2$. Astonishingly, the machine finds that for some values of $0.5 < u^2 < u_{max}^2$, the distance from the local set is \emph{even larger} than in the provenly nonlocal regime. This is a somewhat surprising finding, as one might naively assume that between $0.5$ and $u_{max}^2$ distributions are local, especially when one looks at the nonlocality proof used in the other regime. However, this is not what the machine finds. Instead it gives us a nontrivial conjecture about nonlocality in a new range of parameters $u^2$. Though extracting precise boundaries in terms of $u^2$ for the new nonlocal regime would be difficult from the results in Fig.~\ref{fig:Renou} alone, they strongly suggest that there is some nonlocality in this regime.

%To further strengthen our confidence in this conjecture, in \textbf{Appendix something} we conduct further analysis of the scenario by examining the equivalent problem of deciding the nonlocality of pairs of binary output distributions in the triangle. In this formulation the machine is able to recover the local model at $u_{max}^2$ more precisely, and we reaffirm our numerical evidence in the nonlocality of distributions for $u^2$ between $0.5$ and $u_{max}^2$.
Finally, we have a look at the noise robustness of the distribution with $u^2 = 0.85$, which is approximately the most distant distribution from the local set, from within the provenly nonlocal regime. For the detector efficiency and visibility noise models we recover $\hat{v}^*\approx0.91$, $\hat{v}^*\approx0.89$ respectively, and $\hat{\theta}\approx6^\circ$ for both. Note that these estimates are much more crude than those obtained for the Elegant distributions, primarily due to the target distributions being so much closer to the local set and the neural network getting stuck in local optima. This increases the variations in independent runs of the learning algorithm. E.g. in panel (a) of Fig.~\ref{fig:Renou}, at $u^2=0.85$ the distance is about $0.0034$, whereas in panel (b), in an independent run, the distance for this same point ($v=1$) is around $0.0055$. The absolute difference is small, however the relative changes can have an impact in extracting noise thresholds. Given that the local set is so close to the target distributions (exemplified in the inset in Fig.~\ref{fig:Renou}), it is easily possible that the noise tolerance is smaller than that obtained here.

\section{Discussion}
Let us contrast the presented method to standard numerical techniques. The standard method for tackling the membership problem in network nonlocality is numerical optimization. For a fixed number of possible outputs per party, $o$, without loss of generality one can take the hidden variables to be discrete with a finite alphabet size, and the response functions to be deterministic. In fact the cardinality of the hidden variables can be upper bounded as a function of $o$~\cite{rosset_universal_2017}. Specifically for the triangle this upper bound is $o^3 - o$. This results in a straightforward optimization over the probabilities of each hidden variable symbol and the deterministic responses of the observers, giving $3 (o^3-o -1)$ continuous parameters and a discrete configuration space of size $12(o^3-o)^2$ to optimize over jointly. Note that this is a non-convex optimization space, making it a terribly difficult task. For binary outputs, i.e. $o=2$, this means only 15 continuous variables and a discrete configuration space of 432 possibilities, and is feasible. However, already for the case of quaternary outputs, $o=4$, this optimization is a computational nightmare on standard CPUs with a looming 177 continuous parameters and a discrete configuration space of size 43200. Even when constraining the response functions to be the same for the three parties, $p_A = p_B =p_C$, and the latent variables to have the same distributions, $p_\alpha=p_\beta = p_\gamma$, the problem becomes intractable around a hidden variable cardinality of $8$, which is still much lower than the current upper bound of $60$ that needs to be examined. Standard numerical optimization tools quickly become infeasible even for the triangle configuration - not to mention larger networks!

The causal modeling and Bayesian network communities examine scenarios similar to those relevant for quantum information~\cite{pearl_causality_2000, koller_probabilistic_2009}. The core of both lines of research are directed acyclic graphs and probability distributions generated by them. In these communities there exist methods for this so-called `structure recovery' or `structure learning' task. However, these methods are either not applicable to our particular scenarios or are also approximate learning methods which make many assumptions on the hidden variables, including that the hidden variables are discrete. Hence, even if these learning methods are quicker than standard optimization for current scenarios of interest, they will run into the scaling problem of the latent variable cardinality.

The method demonstrated in this paper attacks the problem from a different angle. It relaxes both the discrete hidden variable and deterministic response function assumptions which are made by the methods previously mentioned. The complexity of the problem now boils down to the response function of the observers - each of which is represented by a feedforward neural network. Though our method is an approximate one, one can increase its precision by increasing the size of the neural network, the number of samples we sum over ($N_{batch}$) and the amount of time provided for learning. Due to universal approximation theorems we are guaranteed to be able to represent essentially any function with arbitrary precision~\cite{cybenko_approximation_1989, hornik_approximation_1991, lu_expressive_2017}. For the first two distributions examined here we find that there is no significant change in the learned distributions after increasing the neural network's width and depth above some moderate level, i.e. we have reached a plateau in performance. Regarding the Elegant distribution, for example, we used depth 5 and width 30 per party. However, we did not do a rigorous analysis in the minimum required size, perhaps an even smaller network would have worked. We were satisfied with the current complexity, since getting a local model for a single target distribution takes a few minutes on a standard computer, using a mini-batch size of $N_{batch}\approx 8000$. For the Renou \emph{et al.} distribution there is still space for improvement in terms of the neural network architecture and the training procedure. The question of what the minimal required complexity of the response functions for a given target distribution is in itself interesting enough for a separate study, and can become a tedious task since the amount of time that the machine needs to learn typically increases with network size.
%In general, studying the ideal hyperparamters for complicated target thoroughly can be tedious since the amount of time that the machine needs to learn typically increases with network size.
%We note, however, that we did not do a rigorous examination of how much this can be reduced while still detecting the same thresholds. Also, for larger network sizes the machine could in principle learn more complex functions. 

We have demonstrated how, by adding noise to a distribution and examining a family of distributions with the neural network, we can deduce information about the membership problem. For a single target distribution the machine finds only an upper bound to the distance from the local set. By examining families of target distributions, however, we get a robust signature of nonlocality due to the clear transitions in the distance function, which match very well with the approximately expected distances.

\section{Conclusion}
In conclusion, we provide a method for testing whether a distribution is classically reproducible over a directed acyclic graph, relying on a fundamental connection to neural networks. The simple, yet effective method can be used for arbitrary causal structures, even in cases where current analytic tools are unavailable and numerical methods are futile, allowing quantum information scientist to test their conjectured quantum, or post-quantum, distributions to see whether they are locally reproducible or not, hopefully paving the way to a deeper understanding of quantum nonlocality in networks.

To illustrate the relevance of the method, we have applied it to two open problems, giving firm numerical evidence that the Elegant distribution is nonlocal on the triangle network, and getting estimates for the noise robustness of both the Elegant and the Renou \emph{et al.} distribution, under physically relevant noise models. Additionally, we conjecture nonlocality in a surprising range of the Renou \emph{et al.} distribution. Our work motivates finding proofs of the nonlocality for both these distributions.

The obtained results on nonlocality are insightful and convincing, but are nonetheless only numerical evidence. Examining whether a certificate of nonlocality can be obtained from machine learning techniques would be an interesting further research direction. In particular, it would be fascinating if a machine could derive, or at least give a good guess for a (nonlinear) Bell-type inequality which is violated by the Elegant or Renou \emph{et al.} distribution. In general, seeing what insight can be gained about the boundary of the local set from machine learning would be interesting. Perhaps a step in this direction would be to understand better what the machine learned, for example by somehow extracting an interpretable model from the neural network analytically, instead of by sampling from it. A different direction for further research would be to apply similar ideas to networks with quantum sources, allowing a machine to learn quantum strategies for some target distributions.

\section{Code Availability}
\label{sec:code}
Our implementation of the method for the triangle network and for the two-party Bell scenario can be found at \url{www.github.com/tkrivachy/neural-network-for-nonlocality-in-networks}.

\begin{acknowledgments}
The authors thank Raban Iten, Tony Metger, Elisa B\"aumer, Marc-Olivier Renou and
Askery Canabarro for fruitful discussions. TK, YC, NG and NB acknowledge financial support from the Swiss National Science Foundation (Starting grant DIAQ, and QSIT), and the European Research Council (ERC MEC). DC acknowledges support from the Ramon y Cajal fellowship, Spanish MINECO (QIBEQI, Project No.\ FIS2016-80773-P, and Severo Ochoa SEV-2015-0522) and Fundaci\'o Cellex, Generalitat de Catalunya (SGR875 and CERCA Program). AT acknowledges financial support
from the UK Engineering and Physical Sciences Research Council (EPSRC DTP).
\end{acknowledgments}
%\bibliography{NN_triangle}

\end{document}